\documentclass[10pt]{article}

\usepackage{ifthen}

\begin{document}

\title{Interaction Of Electrons With Spin Waves In The Bulk And In Multilayers}

\author{L. BERGER} 

\date{Physics Department, Carnegie Mellon University, Pittsburgh, PA 15213}

\maketitle

\vspace{1ex}

\setlength{\baselineskip}{4mm}

The exchange interaction between electrons and magnetic spins is considerably enhanced near interfaces, in magnetic multilayers. As a result, a dc current can be used to generate spin oscillations. We review theory and experimental evidence. The s-d exchange interaction causes a rapid precession of itinerant conduction-electron spins $\vec{s}$ around the localized spins $\vec{S}$ of magnetic electrons. This $\vec{s}$ precession has been observed directly (Weber et al., 2001) with electron beams through Fe, Co, and Ni films. Because of it, the time-averaged interaction torque between $\vec{s}$ and $\vec{S}$ vanishes. Thus, electrons do not interact at all with long-wavelength spin waves, in the bulk. An interface between a magnetic layer and a spacer causes a local coherence between the precession phases of different electrons, in a region within 10 nm from the interface (Slonczewski, 1996; Berger, 1996). Also, a second magnetic layer with pinned $\vec{S}$ is used to "prepare" $\vec{s}$ in a specific direction, before the electrons cross over to the active magnetic layer.  The current-induced drive torque of $\vec{s}$ on $\vec{S}$ in the active layer may be calculated from the spin current (Slonczewski) or from the spin imbalance $\Delta\mu$ (Berger). This torque is equivalent to negative Gilbert damping, leading to instability of $\vec{S}$. Spin current and $\Delta\mu$ are proportional to each other (Berger, 2001) and can arise fom Fermi-surface translation, as well as from expansion/contraction. In fields $\vec{H}$ normal to layers, the critical current $I_{c}$ for $\vec{S}$ instability is predicted to be proportional to the ferromagnetic-resonance frequency $\omega$ (consistent with the Tsoi et al. 1998, 2000, experiments). However, for in-plane $\vec{H}$, due to elliptic $\vec{S}$ precession, $I_{c}$ is not proportional to $\omega$, but linear in H (Katine et al. 2000, experiments). Apart from the current-induced drive torque, an extra Gilbert damping is predicted near the interface even at zero current (Berger, 1996). It has been observed by ferromagnetic resonance (Urban et al., to be published).

\vspace{2ex}

 I. INTRODUCTION AND INTERACTIONS IN THE BULK

\vspace{1ex}

Recently, it has been suggested that the exchange interaction between electrons and magnetic spins is considerably enhanced near interfaces, in magnetic multilayers. As a result, a dc current can be used to generate spin oscillations, or even to switch the spin direction from up to down. Using classical spin models and simple mathematics, we start by describing the situation in bulk samples and single thin films. The effect of interfaces will be discussed in Sections II and III. Then, some of the existing experimental evidence will be covered. The analogy with a semiconductor laser is explained in Section VIII.

In metallic ferromagnets, the s-d exchange interaction $-2J_{sd}\vec{s}\cdot\vec{S}(\vec{r})$ couples $^{1}$ the localized spin $\vec{S}(\vec{r})$ of 3d magnetic electrons to the itinerant spins $\vec{s}$ of 4s conduction electrons. This interaction can be approximately represented by a classical field $\vec{H}_{sd}=-J_{sd}<\vec{S}(\vec{r})>/\mu_{B}$ acting on $\vec{s}$. Here, $\mu_{B}$ is a Bohr magneton, and we have $J_{sd}\simeq 0.1\ eV$ and $\mu_{0}H_{sd}\simeq 1000\ T$.

We assume that a spin wave is present, i.e., that $\vec{S}$ is precessing around the field $\vec{H}$ (Fig. 1a), which includes the external field as well as demagnetizing  and anisotropy fields. In the 1950s and 1960s, theoretical investigations of electron scattering by spin waves were often based on the first Born approximation. In that approximation, s-d exchange is assumed to be a small perturbation. In classical language, this means that $\vec{s}$ remains nearly fixed and nearly parallel or antiparallel to $\vec{H}$ (Fig. 1a), as $\vec{S}$ precesses around $\vec{H}$.
 
However, since $H_{sd}\gg H$, s-d exchange is clearly too large for the Born approximation to be valid. A more correct picture $^{2}$ (Fig. 1b) has $\vec{s}$ precessing rapidly around the instantaneous $\vec{S}$ direction, as $\vec{S}$ itself precesses more slowly around $\vec{H}$. Depending on initial conditions, in many cases $\vec{s}$ follows closely the instantaneous $\vec{S}$ direction, namely,  $\vec{s}$ is nearly parallel to $\vec{S}$ for a spin-up electron (Fig. 1b) and antiparallel for spin-down. This could be called the strong-coupling (or adiabatic) approximation. Because the angle between $\vec{s}$ and $\vec{S}$ is now smaller (Fig. 1b), the instantaneous value of the exchange torque between them is much reduced. Also, because of the rapid $\vec{s}$ precession around $\vec{S}$, that torque changes sign frequently. Finally, different conduction electrons precess out of phase with each other in a bulk sample, so that their torques tend to cancel. We conclude $^{2}$ that electrons do not interact practically at all with spin waves through ordinary s-d exchange, at least for long-wavelength spin waves! 

This interesting conclusion must be modified $^{2}$ if the electron collision rate with solutes or phonons is high and comparable to the $\vec{s}$ precession frequency $\omega_{s}$. Also, if the spin-wave wavelength is short and comparable to $v_{F}/\omega_{s}$, where $v_{F}$ is the electron Fermi velocity; indeed, a current-induced exchange torque leading to a spin-wave instability has been derived $^{3}$ quantum-mechanically for finite spin-wave wavelength. The present paper will emphasize the limit of long wavelength (i.e., uniform $\vec{S}$ precession) and long electron collision time.

The precession of $\vec{s}$ around $\vec{S}$ has been observed directly $^{4}$ by passing a polarized electron beam through a film of Fe, Co or Ni of thickness 0-10 nm (Fig. 1c). The magnetic film is supported by a gold foil of thickness 20 nm. The electron beam is generated by a gun having a GaAs photocathode. The electron spin $\vec{s}$ is originally perpendicular to the fixed magnetic spins $\vec{S}$ of the uniformly magnetized film. After passing through the film, $\vec{s}$ is measured, and is found to be rotated by an angle $\epsilon$ around $\vec{S}$. The value of $\epsilon$ is proportional to magnetic-film thickness. In cobalt, for example, the proportionality constant is $19^{\circ}/nm$. In these experiments, the incident electron energy exceeds the Fermi level in the film by only $\simeq 4-7\ eV$, and its exact value is found to have little effect on $\epsilon$.

\vspace{1ex}

II. EFFECT OF REFERENCE LAYER AND INTERFACE

\vspace{1ex}

S-d exchange torques exerted by conduction electrons on precessing magnetic spins $\vec{S}_{2}$ in layer $F_{2}$ (Fig. 2a) can be increased $^{5,6}$ over their bulk value, by introducing another magnetic layer $F_{1}$ separated from $F_{2}$ by a nonmagnetic spacer N. The magnetic spins $\vec{S}_{1}$ in $F_{1}$ are assumed to have a fixed direction parallel to the precession axis of $\vec{S}_{2}$. If the $F_{1}$ thickness is at least a local spin-diffusion length, this layer is thick enough to control the ratio of spin-up current to spin-down current in the multilayer, and to give to this ratio a value as different from one as possible (as needed $^{6}$ for current-driven experiments). As an example, the spin-diffusion length is 44 nm in Co nanowires $^{7}$ at 77 K. A second role of layer $F_{1}$ is to relax the spin $\vec{s}$ of electrons in $F_{1}$ to a direction parallel (or antiparallel) to $\vec{S}_{1}$, under the combined influence of s-d exchange, spin-orbit interaction and random collisions. Thus, $F_{1}$ prepares $\vec{s}$ in a fixed direction, and could be called the ``reference'' layer. 

Some electrons originating in $F_{1}$ cross over to $F_{2}$ (Fig. 2a). By solving the Schroedinger equation $^{5,6,8}$, while assuming static and uniform $\vec{S}_{2}$, one finds that the $\vec{s}$ direction changes somewhat as the electron crosses the interfaces between these layers, but remains roughly parallel or antiparallel to $\vec{S}_{1}$. Thus, the angle between this $\vec{s}$ and $\vec{S}_{2}$ in $F_{2}$ is large, and comparable to the $\vec{S}_{2}$ precession-cone angle $\theta$ (Fig. 2a). In turn, this implies a large instantaneous torque between $\vec{s}$ and $\vec{S}_{2}$, as in the case of the Born approximation of Fig. 1a. But, as the electron propagates inside $F_{2}$, away from the $N-F_{2}$ interface, $\vec{s}$ precesses rapidly around $\vec{S}_{2}$ (Fig. 2b). As a result, the total torque exerted on all magnetic spins in $F_{2}$, and obtained by summation over all $\vec{s}$ locations in $F_{2}$, is not in the tangential direction shown as $\vec{\tau}_{n}$ in Fig. 2a, as one might naively think. Rather, it is $^{5,6}$ in a direction $\vec{\tau}_{d}$ radial to the $\vec{S}_{2}$ precession orbit; one exception being the case of a very thin layer $F_{2}$ discussed in Section VI.

Electrons with opposite $\vec{s}$ produce opposite torques. Hence, after summation over all electrons, the total torque $\vec{\tau}_{d}$ is found to be zero in equilibrium. An electric current I normal to layers is needed to populate certain electron states near the Fermi level and to depopulate others, resulting in a torque proportional to I. It is also approximately proportional $^{5,6}$ to $sin\theta$ for small $\theta$, and may therefore be written $^{5}$ in the convenient form

\begin{equation}
 \vec{\tau}_{d}=-g(\theta)(\hbar/2eS_{1}S^{2}_{2})I\vec{S}_{2}\times(\vec{S}_{1}\times\vec{S}_{2}),
\end{equation} 

where $g(\theta)$ is a slowly varying function of $\theta$. Note that the direction of a Gilbert or Landau-Lifshitz damping torque is also radial $^{9}$. However, it is directed toward the precession axis of $\vec{S}_{2}$, while the current-induced drive torque $\vec{\tau}_{d}$ is directed outwards provided I flows (Fig. 2a) from $F_{1}$ toward $F_{2}$, i.e., $I>0$.

For two electrons moving in different directions inside $F_{2}$ (Fig. 2b), the phases of $\vec{s}$ precession are equal at first, but become different at a given distance x from the $N-F_{2}$ interface, as x increases from zero. Because of boundary conditions at the interface, the phase for an electron moving nearly parallel to the x axis changes with x at a slower rate than the phase for an electron going in an oblique direction, assuming a spherical Fermi surface. Because of this dephasing effect, the torques exerted on $\vec{S}_{2}$ by different electrons tend to cancel $^{5}$ for x larger than a certain characteristic length $L_{0}=\pi/(k_{\uparrow}-k_{\downarrow})$, where $k_{\uparrow},k_{\downarrow}$ are the spin-up and spin-down wavenumbers in $F_{2}$ (Fig. 2b). From the data of Ref. 4, and assuming a reasonable free-electron Fermi velocity of $1\times 10^{6}\ m/s$, we derive $L_{0}\simeq 9.5\ nm$ and $H_{sd}\simeq 1900\ T$ for cobalt. The length $L_{0}$ plays in the present problem the same role as the transverse spin relaxation time $T_{2}$ in nuclear magnetic resonance. 

Since regions of $F_{2}$ with $x>L_{0}$ do not contribute to the total torque on $F_{2}$ (except $^{8}$ for some periodic fluctuations), the current-induced torque per unit volume of $F_{2}$ is inversely proportional $^{5,6,8}$ to the $F_{2}$ thickness $L^{x}_{2}$ if $L^{x}_{2}>L_{0}$. Electron scattering by solutes or phonons also reduces $^{6}$ the contribution of regions with large x to the torque, and has been observed $^{4}$ to change $\vec{s}$ in the transmitted beam to a direction making a smaller angle with $\vec{S}_{2}$.

\vspace{1ex}

III. SPIN CURRENT AND SPIN IMBALANCE

\vspace{1ex}

In order to calculate the total drive torque $\vec{\tau}_{d}$ on layer $F_{2}$ (Section II), one can introduce $^{5}$ the concept of angular-momentum current $\vec{P}(x)$, often called spin current. It is related to $\vec{\tau}_{d}$ by $\vec{\tau}_{d}=\vec{P}(x=0)-\vec{P}(x=L^{x}_{2})$. In practice, the second term is of no importance if $L^{x}_{2}\gg L_{0}$, for the same reasons (Section II) that regions with $x>L_{0}$ do not contribute to the total torque. This approach has the advantage that the details of the exchange interaction between $\vec{s}$ and $\vec{S}_{2}$ do not appear explicitly in the calculations. Also, it applies even to very large $\theta$ values.   

As an alternative approach, one can introduce $^{6}$ the difference $\Delta\mu =\mu_{\uparrow}-\mu{\downarrow}$ between the Fermi levels $\mu_{\uparrow},\mu_{\downarrow}$ of spin-up and spin-down electrons. In Fig. 2c, we show the occupation numbers $f_{\uparrow}, f_{\downarrow}$ versus energy $\epsilon$ for spin-up and spin-down Fermi distributions at $T\simeq 0$, if $\Delta\mu<0$. Quantum theory $^{9}$ shows that the energy of a spin wave in $F_{2}$ consists of an integer number $n_{m}$ of energy quanta $\hbar\omega$ (magnons), where $\omega$ is the $\vec{S}_{2}$ precession frequency. Note that $n_{m}$ is directly related to the average $\theta$ angle, assumed to be small. The simplest collision process (Fig 3a), caused by s-d exchange and conserving angular momentum, is $^{10}$ such that an electron flips its spin $\vec{s}$ from down to up while one magnon of energy $\hbar\omega$ and wavevector $\vec{q}$ is created. We show a complicated interaction vertex at the center of Fig. 3a, as a reminder that the simple Born approximation (Section I) is not valid for this process. Energy conservation gives $\epsilon_{\downarrow}-\epsilon_{\uparrow}=\hbar\omega$, where $\epsilon_{\uparrow},\epsilon_{\downarrow}$ are the final and initial electron energies in this so-called ``spin-flip process'', also represented by an oblique arrow on Fig. 2c.

Because of the Pauli exclusion principle, such processes must start from a filled state and end in an empty state. Fig. 2c shows that the only suitable initial states for these processes are those located in a region of energy thickness $|\Delta\mu|-\hbar\omega=-(\Delta\mu+\hbar\omega)$ near the top of the spin-down Fermi sea. Thus, the number of possible initial states and, in turn, the resulting drive torque are proportional to $\Delta\mu+\hbar\omega$:  

\begin{equation}
\vec{\tau}_{d}=\alpha_{s}(\theta)(\Delta\mu+\hbar\omega)\frac{\vec{S}_{2}\times(\vec{S}_{1}\times\vec{S}_{2})}{\hbar S_{1}S_{2}}.
\end{equation}

The term $\Delta\mu$ is proportional to the current I, and thus is consistent with Eq.(1). But the new $\hbar\omega$ term represents $^{6}$ a surface damping torque independent of I, with $\alpha_{s}(\theta)$ as the corresponding dimensionless Gilbert parameter. It is found that $\alpha_{s}$ is inversely proportional to $L^{x}_{2}$. This surface damping is a consequence of the Pauli exclusion principle, and is not obtained if the spins $\vec{S}_{2}$ are assumed classical and static as in the Slonczewski theory $^{5}$. Although the present argument leading to Eq. (2) is based on a Fermi distribution with sharp Fermi level at $T\simeq 0$ and with $\Delta\mu+\hbar\omega<0$ (Fig. 2c), Eq. (2) turns out to be valid $^{6}$ even at high temperature $k_{B}T\gg \hbar\omega, \Delta\mu$ where the Fermi level is very blurred, and even when $\Delta\mu+\hbar\omega>0$. 

The current I flowing from $F_{1}$ to $F_{2}$ (Fig. 2a) may act $^{5,6}$ through a translation of the Fermi surfaces in momentum space. In ferromagnets, the spin-up and spin-down currents, as well as the corresponding Fermi-surface displacements (Fig. 3b), are different, leading to non-zero $\vec{P}$ and $\Delta\mu$ at a given point of the Fermi surface, with $|\Delta\mu|\leq 10^{-4}\ eV$. We saw in Section II that the important electrons are those originating in $F_{1}$ and moving toward $F_{2}$. They are located on the $k_{x}>0$ half of the Fermi surface, and indicated by  a filling of dots on Fig. 3b. By summing over all such $k_{x}>0$ states, one can calculate the total spin current $\Sigma\vec{P}(x=0)$ and the average $\Delta\mu$. Actually, only one component of $\vec{P}$, here called $P_{x2}$, is important. It is found $^{11}$ that spin current and $\Delta\mu$ remain in a constant ratio as the current I is varied, or as the ratio of spin-up to spin-down conductivities is varied:

\begin{equation}
\frac{\Sigma P_{x2}(x=0)}{\Delta\mu\cdot sin\theta}=const..
\end{equation}

This shows that the Slonczewski theory $^{5}$, based on $\Sigma P_{x2}$, and the Berger theory $^{6}$, based on $\Delta\mu$, are largely equivalent. The most significant difference between them is that, in their original form, the first one uses phenomenological bulk damping only, and the second surface damping only. This difference has observable consequences, which will be described in Section V and Fig. 4b.

There exists another kind of spin imbalance, associated with isotropic expansion or contraction of the Fermi surfaces (Fig. 3c), which goes back to Aronov and to Johnson and Silsbee $^{12}$. It is the same at all points of the Fermi surface. We denote it by $\overline{\Delta\mu}$, and it can reach $10^{-3}\ eV$. Like $\Delta\mu$, it is related linearly to I, but the relation is more nonlocal. It exists only within a spin-diffusion length from interfaces. As before, only electrons on the $k_{x}>0$ half of the Fermi surface are really active in creating torques, and they are shown by a filling of dots in Fig. 3c. Again, the total spin-current component and average spin imbalance, now denoted by $\Sigma\overline{P}_{x2}$ and $\overline{\Delta\mu}$, can be calculated $^{11}$. Again, we find $\Sigma\overline{P}_{x2}/\overline{\Delta\mu}=const.$, and the constant has nearly the same value as before.

Numerical estimates $^{11}$, based on the assumption of one-dimensional current flow, suggest that $\Sigma\overline{P}_{x2}$ and $\overline{\Delta\mu}$ for Fermi-surface expansion/contraction are each 10 or 100 times larger than the corresponding quantities for translation. Thus, the expansion/contraction mechanism is dominant if the flow is one-dimensional, i.e., if current leads attached to the multilayer sample do not flare out to a larger diameter too close to the multilayer itself.

\vspace{1ex}

IV. CRITICAL CURRENT VERSUS FIELD

\vspace{1ex}

In Section II, we saw that the current-induced drive torque $\vec{\tau}_{d}$ on layer $F_{2}$ (Eq. 1) is directed outwards, i.e., tends to increase $\theta$, if $I>0$. On the other hand, bulk Gilbert damping, characterized by the Gilbert parameter $\alpha_{b}$, gives an inward torque:

\begin{equation}
\vec{\tau}_{b}=-\alpha_{b}\frac{\vec{S}_{2}}{S_{2}}\times\frac{d\vec{S}_{2}}{dt}=\alpha_{b}\omega\frac{\vec{S}_{2}\times(\vec{S}_{1}\times\vec{S}_{2})}{S_{1}S_{2}}.
\end{equation}

Strictly speaking, the second equality only holds for circular precession in a field $\vec{H}$ normal to layers. The physical origin of this bulk damping will be discussed briefly in Section VII.

 At small I or $|\Delta\mu|$, damping dominates, so that $\theta$ actually decreases as time passes (Fig. 3d). However, if I is positive and exceeds a certain critical value $I_{c}$, the drive torque overcomes damping, and $\theta$ increases (Fig. 3e). Depending on the function $g(\theta)$ in Eq. (1), $\theta$ may stabilize at a constant value, or may go all the way to $180^{\circ}$ and stay there. Detailed numerical as well as analytical calculations of dc current-induced precession and switching have been done $^{13}$ for various thin-film geometries and anisotropy cases. And even when the precessing spin distribution in $F_{2}$ becomes non-uniform $^{14}$.

The value of $I_{c}$ is obtained by writing $\vec{\tau}_{d}+\vec{\tau}_{b}=0$, where we take $\vec{\tau}_{d}$ from Eq. (2) rather than Eq. (1) to include surface damping. Using also Eq. (4), and writing $\alpha_{s}=C_{s}/L^{x}_{2}$ and $\Delta\mu=-KI$, where $C_{s}$ and K are positive constants, we obtain

\begin{equation}
I_{c}=\frac{1}{K}(1+\frac{L^{x}_{2}\alpha_{b}}{C_{s}})\hbar\omega.
\end{equation}

In the case of $\vec{H}$ normal to layers, with circular $\vec{S}_{2}$ precession, we have $^{9}$ a frequency $\omega=\gamma(H_{z}-M_{s})$ for the uniform mode. Here, $M_{s}$ is the saturation magnetization in SI units. There is an additional $4\pi$ factor in front of $M_{s}$ if cgs units are used. Eq. (5) shows that $I_{c}$ itself is proportional to $\omega$:

\begin{equation}
I_{c}\propto\omega=\gamma(H_{z}-M_{s}).
\end{equation}

Here, $\gamma$ is the gyromagnetic ratio. Tsoi et al. $^{15,18}$ measured $I_{c}$ on a multilayer consisting of many alternating Co and Cu layers, normal to the field. A sharp silver point (Fig. 4a) contacting the sample was used to achieve large local current densities. Consistent with Eq. (6), they found $I_{c}$ to increase linearly with increasing $H_{z}$, though not quite with the predicted vertical intercept.

The case $^{9}$ of in-plane field $\vec{H}$ with $\omega=\gamma(H_{z}(H_{z}+M_{s}))^{1/2}$ is more common, as it does not require large fields to saturate the sample, but more complicated. Since $\vec{S}_{2}$ precession is now elliptical (Fig. 3f), the bulk damping torque $\vec{\tau}_{b}$ (Eq. 4) and the two parts of $\vec{\tau}_{d}$ vary through a precession cycle, in different manners. As a result, $I_{c}$ is not simply proportional to frequency $\omega$, but rather $^{16}$ linear in $H_{z}$:

\begin{equation}
I_{c}\propto H_{z}+(M_{s}/2).
\end{equation}

Again, an extra $4\pi$ factor should be put in front of $M_{s}$ if cgs units are used. The prediction of Eq. (7) is more or less consistent with $I_{c}$ measurements by Katine et al. $^{16}$ and Albert et al. $^{17}$ on Co/Cu/Co multilayers in in-plane fields, where $I_{c}$ is found to increase linearly with increasing $H_{z}$, with a positive vertical intercept.
         
Tsoi et al. $^{18}$ have succeeded in measuring the frequency $\omega$ of current-induced spin oscillations. At fixed $H_{z}$, they find that $\omega$ increases with increasing I above $I_{c}$. This is not surprising: Since $I_{c}$ increases with increasing $\omega$ (Eq. (6)), the only spin wave excited at $I\simeq I_{c}$ is the one with lowest $\omega$, i.e., the uniform mode with zero wavenumber q (Fig. 3g). At $I>I_{c}$, other modes with larger $\omega$ and q may appear. Slonczewski $^{19}$ has even proposed that the excited modes at $I\simeq I_{c}$ already have $q>0$, due to the radiation of spin waves away from the point contact; this is rather consistent with the measured $\omega$ at $I=I_{c}$ in Tsoi's Fig. 3.

To measure $\omega$, Tsoi et al. adapted a technique well known to radio engineers under the name of ``synchronous rectification'' or of ``direct conversion''. The sample is exposed to a microwave of known and adjustable frequency $\omega_{ex}$ (Fig. 4a). While varying the dc current I through the point contact, they measured the dc ohmic voltage V between point and sample. Due to nonlinear effects, a glitch appears $^{18}$ in dV/dI whenever $\omega_{ex}$ is equal to the frequency $\omega$ of current-induced spin precession.

\vspace{1ex}

V. CRITICAL CURRENT VERSUS LAYER THICKNESS

\vspace{1ex}

According to Eq.(5), $I_{c}$ is a linear function of $F_{2}$ thickness $L^{x}_{2}$, with positive intercept on the $I_{c}$ axis. This prediction is shown as the upper solid line in Fig. 4b. Actually, it should hold for any direction of $\vec{H}$.

$I_{c}$ has been measured versus $L^{x}_{2}$ by Albert et al. $^{20}$, for in-plane field. They found $I_{c}$ to vary linearly with $L^{x}_{2}$, but with no appreciable vertical intercept. If we were to ignore the surface-damping term $\hbar\omega$ in Eq.(2), the constant first term would disappear from Eq. (5), bringing that equation into better agreement with the experimental findings above. This is shown as the lower solid line in Fig. 4b. 

On the other hand, Urban et al. $^{21}$ have recently measured the Gilbert damping constant versus precessing-layer thickness in very thin Fe/Au/Fe/Au multilayers with single-crystal layers, using a transmission-resonance technique. They find a considerable contribution inversely proportional to thickness, in other words, a surface damping. Conceivably, the discrepancy between the two experiments may come from a non-magnetic ``dead'' atomic layer $^{4}$ reducing the effective $L^{x}_{2}$.

\vspace{1ex}

VI. NEXI TORQUES IN VERY THIN LAYERS.

\vspace{1ex}

We saw in Section II that the total current-induced torque on the spins $\vec{S}_{2}$ of layer $F_{2}$ is usually in the radial direction $\vec{\tau}_{d}$ of the $\vec{S}_{2}$ precession orbit of Fig. 2a. However, for very small $F_{2}$ thickness $L^{x}_{2}<L_{0}\simeq 10\ nm$,  the spin $\vec{s}$ of an electron does not have enough time to precess appreciably around $\vec{S}_{2}$ while crossing $F_{2}$ at the Fermi velocity (Fig 2b). Rather, $\vec{s}$ remains  largely in its initial direction parallel or antiparallel to $\vec{S}_{1}$. Therefore, the total torque is in the tangential direction $\vec{\tau}_{n}$ (Fig. 2a), like the torque of a current-induced exchange field of fixed direction parallel or antiparallel to $\vec{S}_{1}$. This effect was first proposed by Heide et al. $^{22}$, who called it non-equilibrium exchange interaction (NEXI), and expected it to exist at all layer thicknesses.

We have calculated the magnitude of $\vec{\tau}_{n}$ as a function of $L^{x}_{2}$, by the same methods $^{8}$ as for $\vec{\tau}_{d}$. The normalized torque per unit volume of $F_{2}$ is plotted versus $k_{N}L^{x}_{2}$ in Fig. 5, for $k_{\uparrow}/k_{N}=1.5$ and $k_{\downarrow}/k_{N}=1.0$. Here, $k_{N}$ is the Fermi wavenumber in layer N. The corresponding plot for $\vec{\tau}_{d}$ was in Fig. 2 of Ref. 8. For $L^{x}_{2}<L_{0}$, $\vec{\tau}_{n}$ is found to be comparable to $\vec{\tau}_{d}$. However, as expected, $\vec{\tau}_{n}$ vanishes at larger $L^{x}_{2}$ values, except for periodic oscillations around zero. The period of these oscillations is approximately $2L_{0}=2\pi/(k_{\uparrow}-k_{\downarrow})$ on the $L^{x}_{2}$ scale. 

\vspace{1ex}

VII. ANISOTROPIC S-D EXCHANGE IN THE BULK

\vspace{1ex}

We saw in Section I that ordinary s-d exchange is not active in the bulk. Anisotropic s-d exchange is $^{23}$ a combination of spin-orbit interaction and exchange, and does not conserve total angular momentum. It is probably responsible for most of the Gilbert damping (Eq. (4)) observed in bulk and single-film magnetic metals. This theoretical expectation is confirmed by the observation $^{24}$ of an increase of damping in Ni and Co at low temperature, of the kind predicted $^{23}$ on the basis of anisotropic s-d exchange.

\vspace{1ex}

VIII. ANALOGY WITH LASER DIODE

\vspace{1ex}
Because of a rather complete analogy with the semiconductor laser, we called $^{6}$ SWASER (Spin Wave Amplification by Stimulated Emission of Radiation) the current-induced spin instability described in the present paper. Like the rate of creation of photons in a laser, the rate of creation $dn_{m}/dt$ for magnons is $^{6}$ proportional to the number $n_{m}$ of magnons itself, and this is called stimulated emission. By opposition, spontaneous emission is independent of $n_{m}$, and is probably negligible $^{25}$ for magnons whenever $\theta\gg 0.1\ ^{\circ}$. The energy gap $\gamma H_{z}$ in the spin-wave spectrum plays in our theory a role like that of the semiconductor band gap in a laser. Also, the spin imbalance $\Delta\mu$ is the analog of the difference $\mu_{v}-\mu_{c}$ between the Fermi levels $^{26}$ of the valence and conduction bands. The spin-wave damping proportional to $\hbar\omega$ (see Section III) also exists for light waves in a laser $^{26}$. As a result, our condition $^{6}$ $\Delta\mu+\hbar\omega=0$ for the critical current has $^{26}$ a correspondent $\mu_{v}-\mu_{c}+\hbar\omega=0$ for the laser threshold. Finally, the $N-F_{2}$ interface acts as a source of momentum $^{8}$ to bridge the momentum gap $k_{\uparrow}-k_{\downarrow}$ between spin-up and spin-down Fermi surfaces. Similarly, solute atoms and lattice pores are used $^{27}$ to bridge the momentum gap between valence and conduction bands in lasers  made of an  indirect-gap semiconductor such as Si. One difference, however, is that spin waves are far more nonlinear than light waves.

The SWASER can be used $^{17}$ to reliably switch small magnetic elements, leading to possible applications for data storage. Also, it constitutes a very compact, tunable, microwave oscillator.    

\vspace{1ex}

\hspace{8em} REFERENCES

\vspace{1em}

1. C. Herring, in $\underline{Magnetism}$, edited by G.T. Rado and H. Suhl (Academic, New York, 1966), Vol. 4, p. 162.

2. L.L. Hirst, Phys. Rev. $\underline{141}$, 503 (1966); J. I. Kaplan, Phys. Rev. $\underline{143}$, 351 (1966); C. Herring, Phys. Rev. $\underline{85}$, 1003 (1952); $\underline{87}$, 60 (1952). Here, instead of the s-d exchange model, often all electrons are assumed itinerant, but a spin precession still happens in the common exchange field.

3. Ya.B. Bazaliy, B.A. Jones and S.C. Zhang, Phys. Rev. B $\underline{57}$, R3213 (1998). 

4. W. Weber, S. Riesen and H.C. Siegmann, Science $\underline{291}$, 1015 (2001).

5. J.C. Slonczewski, J. Magn. Magn. Mater. $\underline{159}$, L1 (1996).
 
6. L. Berger, Phys. Rev. B $\underline{54}$, 9353 (1996).

7. L. Piraux, S. Dubois and A. Fert, J. Magn. Magn. Mater. $\underline{159}$, L287 (1996).

8. L. Berger, J. Appl. Phys. $\underline{81}$, 4880 (1997).

9. A.H. Morrish, $\underline{The\  Physical\  Principles\ Of\ Magnetism}$ (Wiley, New York, 1965). See p. 298 for magnons, p. 549 for Gilbert damping, and p. 545 for the frequency of the uniform mode.

10. A.H. Mitchell, Phys. Rev. $\underline{105}$, 1439 (1957).

11. L. Berger, J. Appl. Phys. $\underline{89}$, 5521 (2001).

12. A.G. Aronov, JETP Letters, $\underline{24}$, 32 (1976); M. Johnson and R.H. Silsbee, Phys. Rev. Lett. $\underline{55}$, 1790 (1985).

13. J.Z. Sun, Phys. Rev. B $\underline{62}$, 570 (2000); Ya.B. Bazaliy, B.A. Jones and S.-C. Zhang, J. Appl. Phys. $\underline{89}$, 6793 (2001); R.H. Koch and C. Heide, paper DF-01, Joint MMM-Intermag Conference, San Antonio, Texas, January 2001.

14. J. Miltat, G. Albuquerque, A. Thiaville and C. Vouille, J. Appl. Phys. $\underline{89}$, 6982 (2001).

15. M. Tsoi, A.G.M. Jansen, J. Bass, W.-C. Chiang, M. Seck, V. Tsoi and P. Wyder, Phys. Rev. Lett. $\underline{80}$, 4281 (1998); $\underline{81}$, 493(E) (1998).

16. J.A. Katine, F.J. Albert, R.A. Buhrman, E.B. Myers and D.C. Ralph, Phys. Rev. Lett. $\underline{84}$, 3149 (2000);

17. F.J. Albert, J.A. Katine, R.A. Buhrman and D.C. Ralph, Appl. Phys. Lett. $\underline{77}$, 3809 (2000).

18. M. Tsoi, A.G.M. Jansen, J. Bass, W. C. Chiang, V. Tsoi and P. Wyder, Nature $\underline{406}$, 46 (2000). 

19. J.C. Slonczewski, J. Magn. Magn. Mater. $\underline{195}$, L261 (1999).

20. F.J. Albert, R.A. Buhrman, J. Katine and D.C. Ralph, paper ED-14, Joint MMM-Intermag Conference, San Antonio, Texas, January 2001.

21. R. Urban, G. Woltersdorf and B. Heinrich, submitted to Phys. Rev. Lett..

22. C. Heide, P.E. Zilberman and R.J. Elliott, Phys. Rev. B, $\underline{63}$, 64424 (2001).

23. V. Kambersky, Can. J. Phys. $\underline{48}$, 2906 (1970); V. Korenman and R.E. Prange, Phys. Rev. B $\underline{6}$, 2769 (1972); L. Berger, J. Phys. Chem. Phys. Solids $\underline{38}$, 1321 (1977).

24. S.M. Bhagat and P. Lubitz, Phys. Rev. B $\underline{10}$, 179 (1974), see Fig. 7; B. Heinrich, D.J. Meredith and J.F. Cochran, J. Appl. Phys. $\underline{50}$, 7726 (1979).

25. L. Berger, Phys. Rev. B $\underline{59}$, 11465 (1999).

26. M.G.A. Bernard and G. Duraffourg, Phys. Stat. Solidi, $\underline{1}$, 699 (1961). 

27. C.B. Duke and N. Holonyak, Physics Today, December 1973, p. 23; R.T. Collins, P.M. Fauchet and M.A. Tischler, Physics Today, January 1997, p.24. 

\vspace{1ex}

\hspace{3em} FIGURE CAPTIONS

\vspace{1ex}

FIG. 1. a) Spin configuration assumed in the Born approximation, with $\vec{s}$ nearly static and antiparallel to $\vec{H}$. b) More correct picture with $\vec{s}$ precessing rapidly around $\vec{S}_{2}$, and following it closely. c) Observation of $\vec{s}$ precession by an angle $\epsilon$ around a fixed $\vec{S}$, in Fe, Co, Ni films (Ref. 4).

FIG. 2. a) Multilayer with magnetic layers $F_{1}, F_{2}$ and nonmagnetic $N,N_{2}$. The oblique arrow represents an electron crossing from $F_{1}$ to $F_{2}$.  b) Two electrons propagating inside $F_{2}$ in different directions have different $\vec{s}$ precession phases around $\vec{S}_{2}$ at a given $x>0$. c) Fermi distributions $f_{\uparrow}(\epsilon), f_{\downarrow}(\epsilon)$ for spin-up and spin-down electrons of energy $\epsilon$, with different Fermi levels $\mu_{\uparrow},\mu_{\downarrow}$, at $T\simeq 0$. The oblique arrow represents a spin-flip process where a magnon of energy $\hbar\omega$ is emitted.

FIG. 3. a) Spin-flip process where a magnon of energy $\hbar\omega$ and wavevector $\vec{q}$ is created. b) Current-induced translations of spin-up and spin-down Fermi surfaces. c) Current-induced expansion or contraction of spin-up and spin-down Fermi surfaces. d) Precession orbit of spin $\vec{S}_{2}$ if $I<I_{c}$. e) Orbit if $I>I_{c}$. f) Orbit if $I>I_{c}$ and $\vec{H}$ is in-plane. g) Dispersion relation $\omega(q)$ for spin waves in a field $H_{z}$. Horizontal lines show the maximum $\omega$ value of current-induced spin waves.

FIG. 4. a) Multilayer with many magnetic layers and a point contact, used by Tsoi et al. (Refs. 15 and 18). b) The upper solid line represents the predicted dependence of $I_{c}$ on $F_{2}$ thickness $L^{x}_{2}$ if both surface and bulk dampings are introduced. The lower solid line applies if only bulk damping is introduced. The crosses represent schematically the experimental data of Albert et al. (Ref. 20).

FIG. 5. Normalized tangential torque component $\tau_{n}$ per unit volume of $F_{2}$, plotted versus normalized layer thickness $L^{x}_{2}$, for fixed I, $\theta$ and $\omega$. The point corresponding to $L^{x}_{2}=L_{0}$ is marked on the horizontal scale. 
 
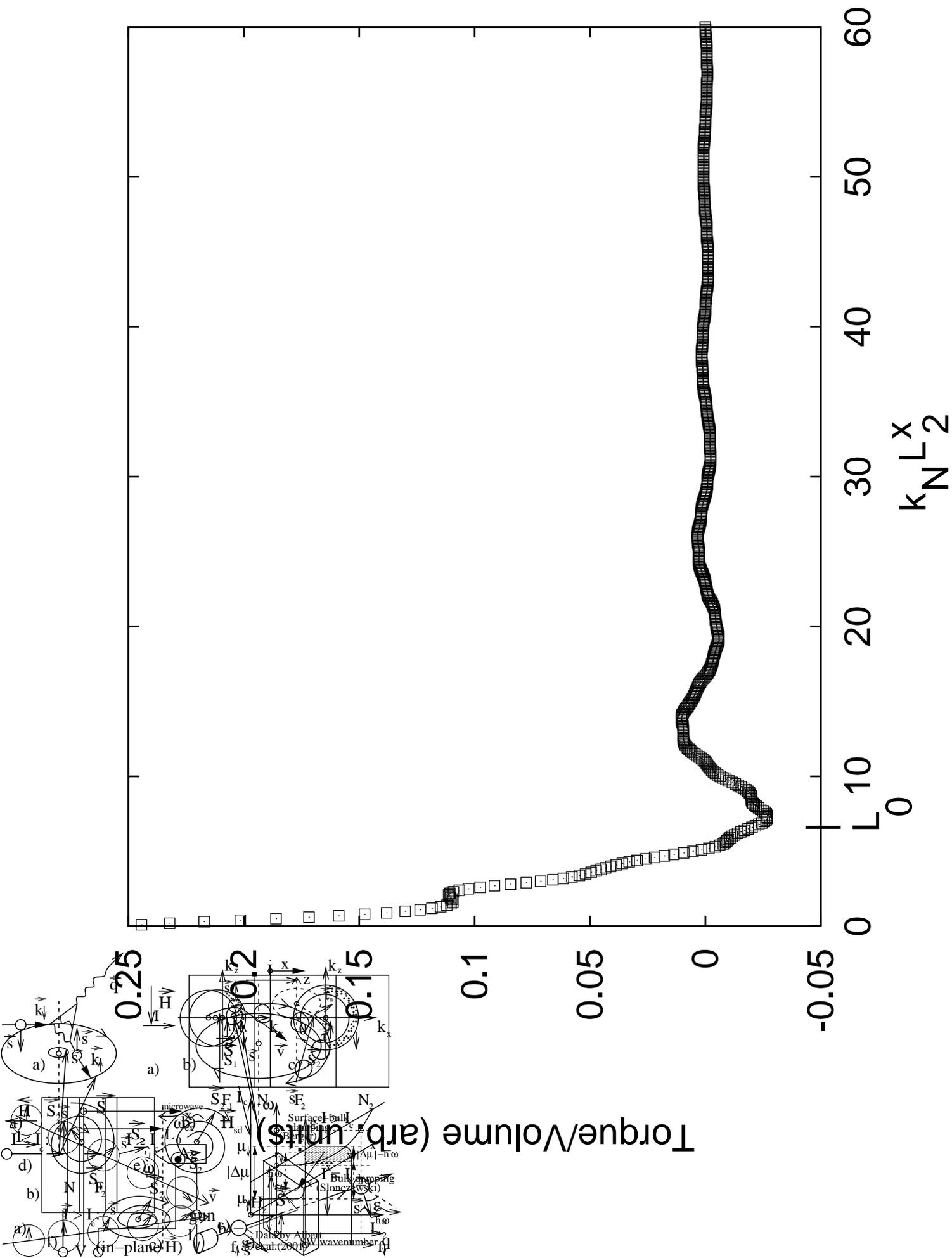
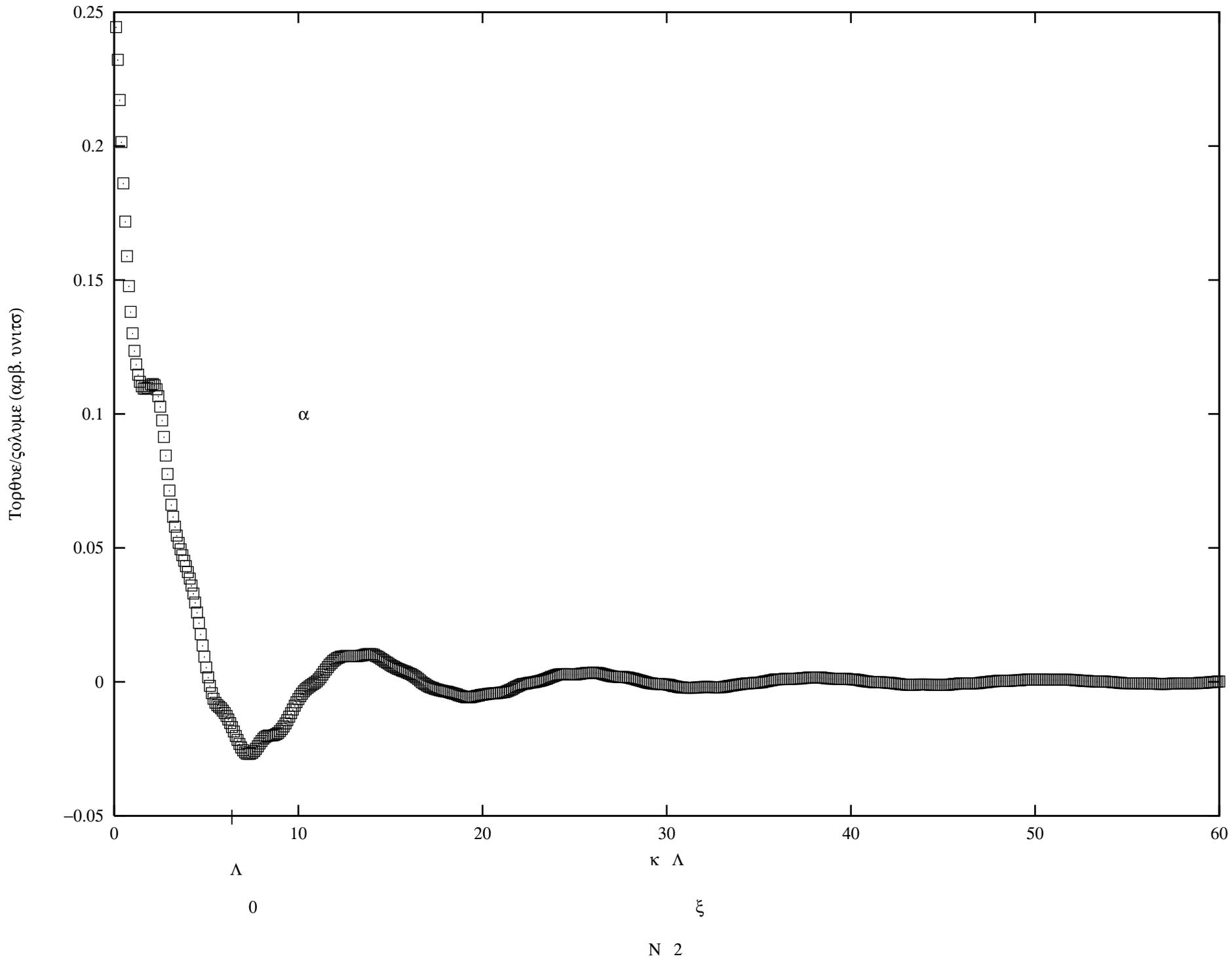
\end{document}